\documentclass[a4paper,10pt,onecolumn,notitlepage,superscriptaddress]{revtex4-1}
\usepackage[latin2]{inputenc}
\usepackage{amsfonts}
\usepackage{amsmath}
\usepackage{amssymb}
\usepackage{amsthm}
\usepackage{subfigure}
\usepackage[T1]{fontenc}
\usepackage{graphicx}
\usepackage{color}
\usepackage{braket}
\usepackage{mathtools}

\usepackage[unicode=true, breaklinks=false, pdfborder={0 0 1}, backref=false, colorlinks=true, linkcolor=blue, citecolor=blue]{hyperref}

\begin{document}

\title[Noisy frequency estimation with noisy probes]{Noisy frequency estimation with noisy probes}

\author{Agnieszka G\'orecka}
\affiliation{School of Physics and Astronomy, Monash University, Clayton, Victoria 3800, Australia}

\author{Felix A. Pollock}
\affiliation{School of Physics and Astronomy, Monash University, Clayton, Victoria 3800, Australia}
	
\author{Pietro Liuzzo-Scorpo}
\affiliation{School of Mathematical Sciences and Centre for the Mathematics and Theoretical Physics of Quantum Non-Equilibrium Systems, University of Nottingham, University Park, Nottingham NG7 2RD, United Kingdom}

\author{Rosanna Nichols}
\affiliation{School of Mathematical Sciences and Centre for the Mathematics and Theoretical Physics of Quantum Non-Equilibrium Systems, University of Nottingham, University Park, Nottingham NG7 2RD, United Kingdom}

\author{Gerardo Adesso}
\affiliation{School of Mathematical Sciences and Centre for the Mathematics and Theoretical Physics of Quantum Non-Equilibrium Systems, University of Nottingham, University Park, Nottingham NG7 2RD, United Kingdom}

\author{Kavan Modi}
\email{kavan.modi@monash.edu}
\affiliation{School of Physics and Astronomy, Monash University, Clayton, Victoria 3800, Australia}
	
\begin{abstract}
We consider frequency estimation in a noisy environment with noisy probes. This builds on previous studies, most of which assume that the initial probe state is pure, while the encoding process is noisy, or that the initial probe state is mixed, while the encoding process is noiseless. Our work is more representative of reality, where noise is unavoidable in both the initial state of the probe and the estimation process itself. We prepare the probe in a GHZ diagonal state, starting from $n+1$ qubits in an arbitrary uncorrelated mixed state, and subject it to parameter encoding under dephasing noise. For this scheme, we derive a simple formula for the (quantum and classical) Fisher information, and show that quantum enhancements do not depend on the initial mixedness of the qubits. That is, we show that the so-called `Zeno' scaling is attainable when the noise present in the encoding process is time inhomogeneous. This scaling does not depend on the mixedness of the initial probe state, and it is retained even for highly mixed states that can never be entangled. We then show that the sensitivity of the probe in our protocol is invariant under permutations of qubits, and monotonic in purity of the initial state of the probe. Finally, we discuss two limiting cases, where purity is either distributed evenly among the probes or concentrated in a single probe.
\end{abstract}
	

\maketitle
\section{Introduction}\label{sec:intro}	

Quantum metrology is a promising research area, where the aim is to develop new quantum technologies that may one day surpass the classical limits of sensing and estimation~\cite{DowlingReview, photonictechnologies, giovannetti2011advances}. Quantum sensing has a wide array of applications, from gravitational wave detection~\cite{PhysRevD.23.1693} to imaging in biological and medical sciences~\cite{bowen}. This is why a great deal of effort has been put into understanding where the power of quantum metrology comes from, but there is no clear answer to this question. Often, the power of quantum metrology is thought to come from quantum entanglement~\cite{pezze}, but entangled states are hard to prepare and inherently fragile~\cite{demkowicz2012elusive}. On the other hand, there are many examples where a quantum enhancement is found even in the absence of quantum entanglement~\cite{Braun}. Here, we add to the latter volume of literature by showing that quantum enhanced scaling is attainable for noisy frequency estimation using a probe in a mixed quantum state.

In the standard case for estimating a parameter $\omega$, that is unitarily encoded, the ultimate limit to sensing is achieved using pure N00N states~\cite{PhysRevLett.85.2733, PhysRevLett.99.163604} (a state that has $n$ photons in superposition in the two arms of an interferometer). Equivalently, for spins and atoms, Greenberger-Horne-Zeilinger (GHZ) states~\cite{PhysRevLett.106.130506} are optimal probe states. These scenarios lead to Heisenberg limited precision of $\Delta \omega \sim1/n$, where $\Delta \omega$ is the standard deviation of the estimate, in contrast to the standard quantum limit (SQL) of $\Delta \omega \sim 1/\sqrt{n}$. The latter is the best precision available to a classical probe. However, this quadratic enhancement quickly disappears with large $n$ when either the initial probe state is noisy~\cite{lloydPRL} or when the encoding is performed in the presence of unavoidable effects of environmental noise~\cite{PhysRevLett.79.3865, escher2011noisy}. Another way to attain a quantum advantage is by extending the length of sensing time~\cite{lloydPRL}. However, this too is limited by the decoherence time scale. Because of this, in the last decade a flurry of research has sought to develop techniques to battle environmental effects~\cite{Sekatskiquantum}.

In practice, preparing N00N, or other highly entangled, states is very difficult and we do not always even have access to pure states. However, the role of entanglement in the initial state of the probe was questioned when quadratic enhancement was observed in a \textit{nuclear magnetic resonance} (NMR) sensing experiment~\cite{Jones2009, PhysRevA.82.022330}, where it is known that there is little or no entanglement present in the quantum state. The observations of these experiments are explained by the calculations of Ref.~\cite{arXiv:1003.1174}, where it is shown that correlations (that may be weaker than entanglement) can lead to a quadratic enhancement in the sensitivity. Expanding on the same ideas, Ref.~\cite{PhysRevA.93.040304} (also see the experimental proposal~\cite{PhysRevA.93.023805}) made use of ideas from the computational model called \textit{deterministic quantum computation with one bit of quantum information} (DQC1)~\cite{KnillLaflamme98} to show that using $n$ maximally mixed qubits, along with one pure qubit, is sufficient to saturate the classical limit to sensing. On the other hand, Refs.~\cite{PhysRevLett.110.240402, PhysRevLett.112.210401} showed that the figure of merit in metrology reduces to a measure of quantum discord~\cite{rmp, discord2} when there is lack of knowledge about the generator of the parameter. These studies have shown that using noisy probe states may not be so bad; however, all of these studies assume that the parameter encoding is a noiseless process.

Recently, several researchers have studied the cases where the parameter encoding is a noisy process. While these studies allow for arbitrary initial states, most often only the optimal performance of the probe is examined in detail, i.e., the initial state of the probe is taken to be pure. Many of these studies are typically concerned with estimating the frequency $\omega$ rather than the phase $\phi$; the two parameters are related to each other by the encoding time $t$, i.e., $\phi = \omega t$. Thus, the total running time of the encoding process itself is regarded as a resource~\cite{PhysRevLett.79.3865}. These studies have shown that a super-extensive growth of the frequency sensitivity may still be attained under time-inhomogeneous, phase-covariant noise \cite{PhysRevA.84.012103, PhysRevLett.109.233601, PhysRevA.92.010102, smirne2015ultimate, Wang2017}, and even more generic Ohmic dissipation \cite{haase2017fundamental}, noise with a particular geometry~\cite{PhysRevLett.111.120401, PhysRevX.5.031010}, or setups related to quantum error correction~\cite{PhysRevLett.112.150802, PhysRevLett.112.080801, PhysRevLett.116.230502}. See also Ref.~\cite{PhysRevA.94.042101,  PhysRevA.95.062307} which question the role of entanglement in such schemes and give advice on practical implementations. In general, these studies have shown that, while the $1/n$ precision scaling in frequency estimation may not be available in the presence of noise, it is possible to achieve a scaling that goes as $1/n^{\frac{3}{4}}$, $1/n^{\frac{5}{6}}$, or $1/n^{\frac{7}{8}}$ depending on the details of the problem. 

In this paper, we combine the tools described in the last two paragraphs. We consider frequency estimation in the presence of dephasing noise, with probe states that are noisy themselves. We describe our setup in Sec.~\ref{sec:initial}, preparation of the probe state in Sec~\ref{sec:prep}, and the noisy process the probe undergoes in Sec.~\ref{sec:enc}. In Secs.~\ref{sec:qfi} and~\ref{sec:cfi} we compute the quantum and classical Fisher information respectively for our protocol; while in Sec.~\ref{sec:uncorr} the Fisher information of the uncorrelated probe is given. In Sec.~\ref{sec:approx}, we demonstrate that the precision of our protocol can be approximately related to a simple function of the purity of the initial probe. Using this we show that, by optimising the protocol, we retain the $1/n^{3/4}$ (or `Zeno scaling') scaling reported in Refs.~\cite{PhysRevA.84.012103, PhysRevLett.109.233601, PhysRevA.92.010102} for any mixedness of the initial probe state. This result shows the robustness of mixed states for the purpose of sensing. Finally, we explore two limiting cases in Sec.~\ref{app:cvc}, and several other properties, such as the permutation symmetry (Sec~\ref{sec:symm}) and monotonicity (Sec.~\ref{sec:mono}) of our protocol. Our conclusions are presented in Sec.~\ref{sec:conclusion}

\section{Initial state and the Protocol}

\begin{figure}[t]
\centering    \includegraphics[width=0.7\textwidth]{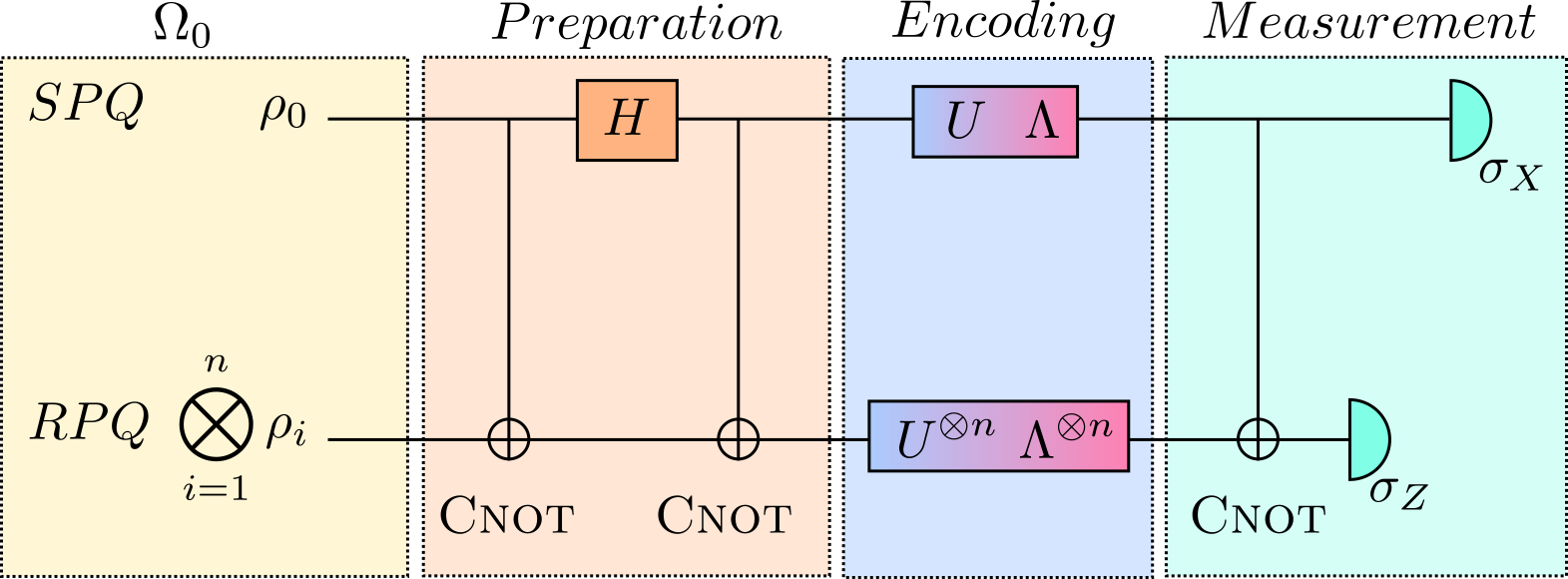}
\caption{\textbf{The protocol.} The initial state is a tensor product of $n+1$ qubits in mixed states $\{ \rho_i \}$. The zeroth qubit is used as a control, and we call it the \emph{Special Probe Qubit} (SPQ). The latter $n$ qubits are \emph{Register Probe Qubits} (RPQ). The protocol has three parts: First the initial state is prepared in a mixed GHZ diagonal state; next, the parameter is encoded in a noisy process; finally, \textsc{Cnot} gates are applied again, before measuring all qubits.}
\label{protocol}
\end{figure}

We begin by laying out the details of the protocol we will consider in this work. First, we describe the initial state of the probe, before discussing the three main stages of the protocol: \textit{preparation}, \textit{parameter encoding} in the presence of noise and, finally, \textit{measurement}. The protocol is graphically illustrated in Fig.~\ref{protocol}.

\subsection{Initial state of the probe}
\label{sec:initial}

Each of the $n+1$ qubits making up the probe can, in general, be in a different mixed state initially, with the $i$th qubit in state $\rho_i$. The overall initial state is the product
\begin{gather}\label{eq:initialstate}
\Omega_0= \bigotimes_{i=0}^n \rho_i, 
\quad\mbox{where}\quad
\rho_{i}=\left( \begin{array}{cc}
  \lambda^{(i)}_{0} & 0 \\
  0 & \lambda^{(i)}_1 \\
 \end{array}
 \right),\quad \lambda^{(i)}_0 = \frac{1+p_{i}}{2} \quad \mbox{and} \quad \lambda^{(i)}_1=\frac{1-p_{i}}{2}.
\end{gather}
Here, $\lambda^{(i)}_0$ and $\lambda^{(i)}_1$ are the probabilities of the $i$th qubit being in state $\ket{0}$ and $\ket{1}$ respectively. This setup could describe an experiment which senses magnetic or gravitational fields using an ensemble of atoms~\cite{mansell, PhysRevA.93.023805}. Such an ensemble can be initially aligned with the direction an external magnetic field so that the Bloch vector of each qubit will be pointing in the same direction. In that case we have $p_{i} \in [0,1]$ and $\lambda^{(i)}_0 \geqslant \lambda^{(i)}_1$.

This initial state may be fully described by a vector $\mathbf{p}=(p_0, p_1, \dots, p_n)$. By setting these parameters to different values, we can obtain any combination of diagonal mixed states ranging from fully mixed ($p_i=0$) to pure ($p_i=1$). In Sec.~\ref{sec:approx} we will show a simple relationship between the sensitivity of our probe and $\mathbf{p}$. And in Sec.~\ref{app:cvc}, we will examine two special cases, which we dub \emph{uniform}, where all $p_i =p$ for $i \in [0,n]$ and \emph{tilted}, where $p_0 = 1$ and $p_i=0$ for all $i>0$.

As the $0$th qubit is used to make control operation on the rest, we call it the \emph{special probe qubit} (SPQ), while the latter $n$ qubits are called \emph{register probe qubits} (RPQs). Throughout this article we represent the state of the RPQs in the matrix basis of the SPQ:
\begin{gather}
\Omega_0= 
 \left( \begin{array}{cc}
  \lambda_0  \bigotimes_{i=1}^n \rho_i & 0 \\
  0 & \lambda_1 \bigotimes_{i=1}^n \rho_i
 \end{array} \right),
\end{gather}
where for simplicity we have dropped the superscript $0$ from the $\lambda$'s in the state $\rho_0$.

\subsection{Preparation of the probe}
\label{sec:prep}

Parameter estimation with a pure state is optimal when the probe state is taken to be a GHZ state $\frac{1}{\sqrt{2}}(\ket{0\dots 0}+\ket{1\dots 1})$. To attain optimal performance in this limit, we adopt a preparation procedure which creates a GHZ state in the case $\mathbf{p}=(1,\dots, 1)$; more generally, we end up with a mixed GHZ-diagonal state, i.e., a mixture of state of the form $\frac{1}{\sqrt{2}} (\ket{\mathbf{k}}+\ket{\mathbf{1} -\mathbf{k}})$, where $\mathbf{k} := (k_{1}, \dots, k_{n})$ with $k_{j}=0,1$. Here the superscript denotes the $j$th qubit, and $\mathbf{1} := (1, \dots, 1)$.

In fact, we conjecture that GHZ diagonal states lead to optimal performance for any configuration of noisy qubits in Eq.~\eqref{eq:initialstate}. The preparation involves three steps: First we apply a set of controlled-not (\textsc{Cnot}) gates to all RPQs with the SPQ as the control. The \textsc{Cnot} gate applies the Pauli matrix $X$ to the target if the control qubit SPQ is in the state $\ket{1}$, and the identity operator $\mathbf{I}$ if the control qubit is in the state $\ket{0}$. Next, a Hadamard gate is applied to the SPQ, followed by another set of \textsc{Cnot} gates on the RPQs, again controlled by the SPQ.

The state after the full preparation procedure has the following form:
\begin{gather}
\Omega_{\rm probe} =
\frac{\lambda_0}{2}\left( \begin{array}{cc}
  \bigotimes_{i=1}^n \rho_i &\bigotimes_{i=1}^n \rho_i X_i\\
  \bigotimes_{i=1}^n X_i \rho_i &  \bigotimes_{i=1}^n X_i \rho_i X_i
 \end{array} \right) \!+\! \frac{\lambda_1}{2}\left( \begin{array}{cc}
  \bigotimes_{i=1}^n X_i \rho_i X_i &  -\bigotimes_{i=1}^n X_i \rho_i  \\
  -\bigotimes_{i=1}^n \rho_i X_i &  \bigotimes_{i=1}^n \rho_i 
 \end{array} \right).\label{omprobe}
\end{gather}
The subscript $i$ in $X_i$ denotes that the $X$ operator acts on the $i$th qubit. Note here that the matrix $X_i \rho_i X_i$ is diagonal in the same basis as $\rho_i$, but with the eigenvalues switched:
\begin{gather}
X_i \rho_{i} X_i=\left( \begin{array}{cc}
  \lambda_1^{(i)} & 0 \\
  0 & \lambda_0^{(i)}
 \end{array} \right).
\end{gather}

This prepared state is a GHZ diagonal state, which is easily seen by tracking how its eigenvectors are built up. After the application of the first \textsc{Cnot} gate and the Hadamard gate, the eigenvalues and eigenvectors of the probe state are
\begin{gather}
\left\{\lambda_0 \lambda^{(1)}_{k_1} \dots \lambda^{(n)}_{k_n}, \ \ket{+, \mathbf{k}} \right\}
\quad \mbox{and} \quad
\left\{\lambda_1 \lambda^{(1)}_{k_1} \dots \lambda^{(n)}_{k_n}, \ \ket{-, \mathbf{1-k}} \right\},
\end{gather}
where $\ket{\pm} = \frac{1}{\sqrt{2}}(\ket{0}\pm\ket{1})$. After the second \textsc{Cnot} gate, the eigenvectors and eigenvalues of the probe state are
\begin{gather}
\left\{g_{\mathbf{k}}^{(+)}, \ \ket{G_{\mathbf{k}}^{(+)}}:=
\frac{\ket{0, \mathbf{k}}+
\ket{1, \mathbf{1-k}}}{\sqrt{2}} \right\},
\quad \left\{g_{\mathbf{1-k}}^{(-)}
, \ \ket{G_{\mathbf{1-k}}^{(-)}}:=
\frac{\ket{0, \mathbf{1-k}}-
\ket{1, \mathbf{k}}}{\sqrt{2}} \right\}, \label{eq:eigenprobe}
\end{gather}
where $g_{\mathbf{k}}^{(+)}=\lambda_0 \lambda^{(1)}_{k_1} \dots \lambda^{(n)}_{k_n}$ and $g_{\mathbf{1-k}}^{(-)} := \lambda_1 \lambda^{(1)}_{k_1} \dots \lambda^{(n)}_{k_n}$. While the eigenvalue corresponding to $\ket{G_{\mathbf{k}}^{(-)}}$ is $g_{\mathbf{k}}^{(-)} = \lambda_1 \lambda^{(1)}_{1-k_1} \dots \lambda^{(n)}_{1-k_n}$, which we will make use of a little later.

The application of two \textsc{Cnot} gates may seem redundant. However, it is well known that this probe outperforms the probe where the first \textsc{Cnot} is omitted~\cite{PhysRevA.82.022330, arXiv:1003.1174}. We will show in Sec.~\ref{sec:symm} that the first \textsc{Cnot} gate also introduces a nice symmetry in our protocol.

\subsection{Parameter encoding in presence of noise}
\label{sec:enc}

We are now ready to encode the parameter we want to estimate onto our prepared state. The parameter is encoded via free evolution of the probe. However, as we encode the desired parameter, the free evolution also introduces noise, which can be described by a phenomenological quantum master equation~\cite{PhysRevA.73.012111, PhysRevA.75.062103, PhysRevA.81.062120}. 

For simplicity, we restrict ourselves to a pure dephasing process that commutes with the encoding process. This means we can treat the parameter encoding and noise as occurring sequentially. With minor modifications we can consider a more general phase-covariant form for the noisy process\footnote{In an adjoining article~\cite{nottinghampaper} we do consider phase-covariant noise, but restrict the probe qubits to be identical initially.}. However, the case we consider is sufficient to show that `Zeno scaling' is possible even when the probe state is noisy.

The parameter $\omega$ stems from a classical field, e.g. magnetic field, and it is applied identically to all qubits through the unitary operation $U^{\otimes {n+1}}$, with $U=\exp[-i\omega t Z/2]$, where $Z$ is a Pauli matrix and $t$ is the time of free evolution. The system just after the free evolution is in the following state:
\begin{align}
 \Omega_{\omega}=&  \frac{\lambda_0}{2} \left( \begin{array}{cc}
  \bigotimes_{i=1}^n \rho_{i} & e^{-i \omega t}\bigotimes_{i=1}^n U_i \rho_{i} X_{i} U_{i}^{\dagger} \\
  e^{i \omega t}\bigotimes_{i=1}^n U_i X_i \rho_i U_i^\dagger &  \bigotimes_{i=1}^n X_i \rho_i X_i 
 \end{array} \right) \nonumber \\
 &+\frac{\lambda_1}{2} \left( \begin{array}{cc}
  \bigotimes_{i=1}^n X_i \rho_i X_i &  -e^{-i \omega t}\bigotimes_{i=1}^n U_i X_i \rho_i U_i^\dagger \\
  -e^{i \omega t}\bigotimes_{i=1}^n U_i \rho_i X_i U_i^\dagger &  \bigotimes_{i=1}^n \rho_i
 \end{array} \right).
\end{align}
The unitary operator on the SPQ has already been applied implicitly. Note that the states $\rho_i$ and $X_i \rho_i X_i$ commute with the unitary matrix $U_i$.  As before, the subscript $i$ in $U_i$ denotes the unitary operator action on the state $\rho_i$.

While the parameter is encoded, the probe is also subjected to dephasing noise. We can describe such a noisy process on the probe by the action of a superoperator $\Lambda$ that consists of two Kraus operators $L_0$ and $L_1$:
\begin{align}
& \Lambda [\rho] = \sum_{i=0,1} L_i \rho L_i^\dagger, \nonumber\\\label{lambda}
&\mbox{where}  \quad  L_0 = \sqrt{\frac{1+\exp(- g t^{\alpha})}{2}} \mathbf{I} \quad \mbox{and} \quad L_1=\sqrt{\frac{1-\exp(- g t^{\alpha})}{2}} Z,
\end{align}
with the rate of dephasing $g$ a positive real number and $\alpha$ a constant which determines the type of noise we face. When $\alpha=1$, the noisy process is described by a semigroup~\cite{gks1976, lindblad1976}, i.e., $\Lambda_\tau \circ \Lambda_t = \Lambda_{\tau + t}$. On the other hand, $\alpha \ne 1$ corresponds to a time-inhomogeneous indivisible process. 

Although the encoding and the noisy process are happening simultaneously, we can consider them as sequential processes, because the actions of $\Lambda$ and $U$ commute. Using the fact that $\Lambda_i[\rho_i]= \Lambda_i[X_i \rho_i X_i] = \rho_i$, when $\rho_i$ is diagonal, and  $U_i\Lambda_i[\rho]U_i^\dag= \Lambda_i [U_i \rho U_i^\dag]$ for any $\rho$, we can compute the state after the noisy encoding:
\begin{align}
\Omega_{\omega,g} =& \frac{\lambda_0}{2} \left( \begin{array}{cc}
\bigotimes_{i=1}^n \rho_i & e^{-i \omega t} e^{-g t^{\alpha}} \bigotimes_{i=1}^n \Lambda_i[U_i \rho_i X_i U_i^\dagger] \\
e^{i \omega t} e^{-g t^{\alpha}} \bigotimes_{i=1}^n \Lambda_i[U_i X_i \rho_i U_i^\dagger] &  \bigotimes_{i=1}^n X_i \rho_i X_i
\end{array} \right) \nonumber\\
&+\frac{\lambda_1}{2} \left( \begin{array}{cc}
\bigotimes_{i=1}^n X_i \rho_i X_i &  -e^{-i \omega t} e^{-g t^{\alpha}} \bigotimes_{i=1}^n \Lambda_i[U_i X_i \rho_i U_i^\dagger] \\
-e^{i \omega t} e^{-g t^{\alpha}} \bigotimes_{i=1}^n \Lambda_i[U_i \rho_i X_i U_i^\dagger] &  \bigotimes_{i=1}^n \rho_i
\end{array} \right).\label{eq:prepstate}
\end{align}

\subsection{Probe measurement (all qubits)}

The measurement procedure consists of two steps: a final \textsc{Cnot} gate applied to all RPQs with the SPQ as the control followed by a measurement of the RPQs in the $Z$ basis and the SPQ in the $X$ basis. In Appendix~\ref{app:onequbmeas}, we consider a different readout strategy, where only the SPQ is measured and all RPQs are discarded. In the case we are treating here, the final (pre-measurement) state, after the third \textsc{Cnot} gate, takes the following form
\begin{align}
\Omega_{\rm final} \!=& \frac{\lambda_0}{2} \left( \begin{array}{cc}
  \bigotimes_{i=1}^n \rho_i & e^{-i \omega t} e^{-(n+1)g t^{\alpha}} \bigotimes_{i=1}^n U_i^2\rho_i  \\
  e^{i \omega t} e^{-(n+1)g t^{\alpha}} \bigotimes_{i=1}^n \rho_i {U_i^\dagger}^2 &  \bigotimes_{i=1}^n \rho_i 
 \end{array} \right) \\\nonumber
&\!+\! \frac{\lambda_1}{2} \left( \begin{array}{cc}
  \bigotimes_{i=1}^n X_i \rho_i X_i &  \!-e^{-i \omega t} e^{-(n+1)g t^{\alpha}} \bigotimes_{i=1}^n U_i^2 X_i \rho_i X_i\\
  \!-e^{i \omega t} e^{-(n+1)g t^{\alpha}} \bigotimes_{i=1}^n X_i \rho_i X_i {U_i^\dagger}^2 &  \bigotimes_{i=1}^n  X_i \rho_i X_i
 \end{array} \right)\!.
 \label{om1}
\end{align}

Going from Eq.~\eqref{eq:prepstate} to Eq.~\eqref{om1}, we have used several properties of the superoperator $\Lambda$. We will describe the transformation for the $i$th qubit, and will therefore drop the subscript $i$ for the moment. Since the Kraus operator $L_0$ in Eq.~\eqref{lambda}  is proportional to the identity matrix, it commutes with everything. The $L_1$ Kraus operator is proportional to the $Z$ Pauli matrix, therefore $L_1 X =- X L_1$. From this, one can show that the superoperator $\Lambda$ acting on an arbitrary $2 \times 2$ matrix $\eta$ has the following properties: $ \Lambda[\eta] X = L_0 \eta X L_0 - L_1 \eta X L_1$. Setting $\eta=U \rho X U^\dag$, the upper left element of the first matrix of Eq.~\eqref{eq:prepstate} becomes $\Lambda[U \rho X U^\dag]X = L_0 U \rho X U^\dag X L_0 - L_1 U \rho X U^\dag X L_1$. Next, we have that $U=X U^\dag X$ and the fact that $U, \ \rho, \ L_0, \ \mbox{and} \ L_1$ all commute with each other, which yields $\Lambda[U \rho X U^\dag]X = L_0^2 U^2 \rho - L_1^2 U^2 \rho = \exp[-g t^{\alpha}] U^2 \rho$. The derivation of the other off-diagonal terms above follow in the same manner. 

Next, the RPQs are measured in the $Z$ basis. Each measurement can result in a qubit being either in state $\ket{0}_i$ or $\ket{1}_i$, and the given outcome of the measurement is fully defined by a vector $\mathbf{k} = (k_1, \dots, k_n)$, where $k_i=0,1$ corresponds to state $\ket{0}_i$ or $\ket{1}_i$ respectively for the $i$th RPQ. The overall phase factor coming from the action of the $U^2$ operator depends only on the Hamming weight $m_{\mathbf{k}}$ of $\mathbf{k}$. Every time we measure state $\ket{0}_i$, we get a factor of $\exp[-i\omega t]$, while for state $\ket{1}_i$ we get a factor of $\exp[i\omega t]$. Let's assume that for a given permutation of $\mathbf{k}$ we measure $m_{\mathbf{k}}=\sum_{i=1}^n k_i$ times state $\ket{1}$; this means that we measured state $\ket{0}$ $n-m_{\mathbf{k}}$ times. Therefore, we get a phase factor of $\exp[i m_{\mathbf{k}} \omega t] \times \exp[-(n-m_{\mathbf{k}})i \omega t] = \exp[-i(n-2m_{\mathbf{k}}) \omega t]$ for this permutation.

The SPQ state after observing a particular string of RPQ outcomes $\mathbf{k}$ is
\begin{align}
\rho_{SPQ}^{(m_{\mathbf{k}})}
 =&  \frac{\lambda_0}{2}  \left( \begin{array}{cc}
1 & e^{-i (n+1-2m_{\mathbf{k}}) \omega t} e^{-(n+1)g t^{\alpha}} \\
  h.c. & 1
 \end{array} \right) \prod_{i=1}^n \lambda^{(i)}_{k_i}\nonumber\\
&+ \frac{\lambda_1}{2} \left( \begin{array}{cc}
1 &  -e^{-i (n+1-2m_{\mathbf{k}}) \omega t} e^{-(n+1)g t^{\alpha}} \\
  h.c. &  1
 \end{array} \right) \prod_{i=1}^n \lambda^{(i)}_{1-k_i}.
\end{align}
Next, we measure the SPQ in the $X$ basis to get the probabilities $q^{(\pm)}_{m}:= \bra{\pm} \rho_{SPQ}^{(m_{\mathbf{k}})} \ket{\pm}$, which we can expand to get
\begin{align}
q_{m_{\mathbf{k}}}^{(\pm)}=&\frac{\lambda_0}{2} \prod_{i=1}^n \lambda^{(i)}_{k_i} \left\{ 1 \pm e^{-(n+1)g t^{\alpha}} \cos[(n+1-2m_{\mathbf{k}})\omega t]\right\}
\nonumber\\\label{eq:prob}
&+\frac{\lambda_1}{2} \prod_{i=1}^n \lambda^{(i)}_{1-k_i} \left\{ 1 \mp e^{-(n+1)g t^{\alpha}} \cos[(n+1-2m_{\mathbf{k}}) \omega t]\right\}.
\end{align}

In the next section, we will use these probabilities to compute the classical Fisher Information of this distribution. But first, in order to determine the optimality of this measurement scheme, we compute the quantum Fisher Information for our protocol.

\section{Fisher Information}

Let us suppose a total time of $\mathcal{T}$ is allocated to the estimation procedure. In the protocol above, each experiment has a running time of $t$, allowing for the experiment to be repeated $\mathcal{T}/t$ times. The precision in estimating $\omega$ is bounded by the Cram\'{e}r-Rao bound~\cite{cramer, rao}
\begin{gather}
\Delta \omega \geqslant\frac{1}{\sqrt{F_C}} \quad \mbox{where} \quad F_C = \frac{\mathcal{T} \mathcal{F}_C}{t}.
\end{gather}
Here, $\Delta \omega$ is the standard deviation of our estimate of $\omega$, $F_C$ is the total classical Fisher information (CFI) after time $\mathcal{T}$. This quantity is related to $\mathcal{F}_C$, the CFI of a single experiment of length $t$, which is obtained from the measurement statistics given in Eq.~\eqref{eq:prob}.

Finally, the CFI is bounded by the quantum Fisher information (QFI) $\mathcal{F}_C \leqslant \mathcal{F}_Q$, where the latter may be derived by optimising the CFI over all measurement strategies~\cite{CavesBraunstein94}. The CFI can saturate the QFI~\cite{Cover}, provided it is possible to implement a suitable measurement. Thus the QFI points to the ultimate precision that could be achieved with our probe state. However, in practice, the required measurements may be non-trivial~\cite{micadei} and our particular choice of measurement may lead to a gap between the CFI and QFI. Thus, we first compute the QFI and then show that our measurement scheme indeed leads to a CFI that saturates the corresponding QFI.

\subsection{Quantum Fisher Information (QFI)}
\label{sec:qfi}

To compute the QFI, we write the probe state in Eq.~\eqref{omprobe} in its eigenbasis $\Omega_{\rm probe} = \sum_{\mathbf{k},\pm} g_{\mathbf{k}}^{(\pm)} \ket{G_{\mathbf{k}}^{(\pm)}} \bra {G_{\mathbf{k}}^{(\pm)}}$. The eigenvectors and eigenvalues of our probe are given in Eq.~\eqref{eq:eigenprobe}. If our encoding process were noiseless, we could compute the QFI using the formula~\cite{CavesBraunstein94, Braunstein96}
\begin{gather}\label{eq:qfi}
\mathcal{F}_Q = 2 \sum_{\mathbf{k}, \mathbf{k}'} \sum_{r',r\in\{\pm\}} \frac{\left(g_{\mathbf{k}}^{(r)}-g_{\mathbf{k'}}^{(r')}\right)^2}{g_{\mathbf{k}}^{(r)}+g_{\mathbf{k'}}^{(r')}} |\braket{G_{\mathbf{k}}^{(r)}|\mathcal{G}|G_{\mathbf{k'}}^{(r')}}|^2,
\end{gather}
where $\mathcal{G}$ is the generator encoding the parameter. Unfortunately, our process is not noiseless. However, since the noisy process and the unitary process commute, we can first subject the probe to noise to arrive at $\tilde{\Omega}_{\rm probe} := \Lambda^{\otimes n+1}[\Omega_{\rm probe}]$, before using Eq.~(\ref{eq:qfi}), provided we know the eigen-decomposition of $\tilde{\Omega}_{\rm probe}$.

Let's consider the action of the noise on the operator $\Gamma_{\mathbf{k}}= g_{\mathbf{k}}^{(+)} \ket{G_{\mathbf{k}}^{(+)}} \bra {G_{\mathbf{k}}^{(+)}} + g_{\mathbf{k}}^{(-)} \ket{G_{\mathbf{k}}^{(-)}} \bra {G_{\mathbf{k}}^{(-)}}$ for some choice of $\mathbf{k}$. We expand the states in the computational basis to get
\begin{align}
\Lambda^{\otimes n+1}[\Gamma_{\mathbf{k}}] =& \left(g_{\mathbf{k}}^{(+)}+g_{\mathbf{k}}^{(-)} \right) \Lambda^{\otimes n+1}[\ket{0, \mathbf{k}}\bra{0, \mathbf{k}}+\ket{1, \mathbf{1-k}}\bra{1, \mathbf{1-k}}] \\\nonumber
&+ \left(g_{\mathbf{k}}^{(+)}-g_{\mathbf{k}}^{(-)} \right) \Lambda^{\otimes n+1}[\ket{0, \mathbf{k}}\bra{1, \mathbf{1-k}}+\ket{1, \mathbf{1-k}}\bra{0, \mathbf{k}}] \\
=& \tilde{g}_{\mathbf{k}}^{(+)} \ket{G_{\mathbf{k}}^{(+)}} \bra {G_{\mathbf{k}}^{(+)}} + \tilde{g}_{\mathbf{k}}^{(-)} \ket{G_{\mathbf{k}}^{(-)}} \bra {G_{\mathbf{k}}^{(-)}},
\end{align}
where we have used $\Lambda^{\otimes n+1}[\ket{0, \mathbf{k}}\bra{1, \mathbf{1-k}} = e^{-(n+1)g t^{\alpha}} [\ket{0, \mathbf{k}}\bra{1, \mathbf{1-k}}$. In other words, the noise maps the GHZ diagonal state into another GHZ diagonal state with new eigenvalues
\begin{gather}
\tilde{g}_{\mathbf{k}}^{(\pm)} = \frac{{g}_{\mathbf{k}}^{(+)}+{g}_{\mathbf{k}}^{(-)}}{2} \pm e^{-(n+1)g t^{\alpha}} \frac{{g}_{\mathbf{k}}^{(+)}-{g}_{\mathbf{k}}^{(-)}}{2}.
\end{gather}

To compute the QFI, we write down the generator $\mathcal{G} = \frac{t}{2} \sum_{i} Z^{(i)}  \otimes \mathbf{I}^{(\bar{i})}$. Its action on an eigenvector yields 
\begin{align}
\mathcal{G} \ket{G_{\mathbf{k}}^{(\pm)}} =&
\frac{t}{2\sqrt{2}} [(n+1-2m_{\mathbf{k}}) \ket{0, \mathbf{k}} \pm (m_{\mathbf{k}} -(n-m_{\mathbf{k}})+1) \ket{1, \mathbf{1-k}}] 
\nonumber\\
=& t \frac{n+1-2m_{\mathbf{k}}}{2} \ket{G_{\mathbf{k}}^{(\mp)}},
\end{align}
where again $m_{\mathbf{k}}=\sum_{i=1}^n k_i$. From this we have that $\braket{G_{\mathbf{k'}}^{(r')} | \mathcal{G} | G_{\mathbf{k}}^{(r)}} = t \frac{n+1-2m_{\mathbf{k}}}{2} \delta_{\mathbf{k} \mathbf{k'}} \delta_{r,r'+1}$. Immediately, we find the QFI to be
\begin{align}
\label{QFIeq}
\mathcal{F}_Q =& t^2 \sum_{\mathbf{k}}  \frac{\left(\tilde{g}_{\mathbf{k}}^{(+)} -\tilde{g}_{\mathbf{k}}^{(-)} \right)^2} {\tilde{g}_{\mathbf{k}}^{(+)}+\tilde{g}_{\mathbf{k}}^{(-)}} (n+1-2m_{\mathbf{k}})^2 
= t^2 e^{-2 (n+1)g t^{\alpha}} \sum_{\mathbf{k}} \mathcal{P}_{\mathbf{k}} (n+1-2m_{\mathbf{k}})^2, \\
& \mbox{where} \quad \mathcal{P}_{\mathbf{k}}:=
\frac{\left({g}_{\mathbf{k}}^{(+)} -{g}_{\mathbf{k}}^{(-)} \right)^2} {{g}_{\mathbf{k}}^{(+)}+{g}_{\mathbf{k}}^{(-)}} 
=\frac{\left( \lambda_0 \prod_{i=1}^n \lambda^{(i)}_{k_i}-\lambda_1 \prod_{i=1}^n \lambda^{(i)}_{1-k_i} \right)^2 }
{\lambda_0 \prod_{i=1}^n \lambda^{(i)}_{k_i}+\lambda_1 \prod_{i=1}^n \lambda^{(i)}_{1-k_i}}. \label{eq:Pofk}
\end{align}
We will shortly show that our readout scheme yields the same value for the CFI, i.e., it is experimentally possible to achieve the theoretical maximum precision.

\subsection{Classical Fisher Information (CFI)}
\label{sec:cfi}

The CFI to estimate parameter $x$ is given by the following formula:
\begin{gather}
\label{cfieqgen}
\mathcal{F}_C = \sum_i \frac{(\partial_x q_{i})^2}{q_{i}},
\end{gather}
where the $q_i$ are measurement outcomes, in our case taken from Eq.~\eqref{eq:prob}.

Our readout scheme begins with measuring $m_{\mathbf{k}}$ RPQs in state $\ket{1}$ (and $n-m_{\mathbf{k}}$ in state $\ket{0}$), followed by a measurement of the SPQ in the $\ket{\pm}$ basis. Since different qubits have different purity, i.e., different values of $p_i$, the CFI depends on the specific string of outcomes: $\mathcal{F}_{C} = \sum_{\mathbf{k}} f_{\mathbf{k}}$ with
\begin{gather}
\label{smallf}
f_{\mathbf{k}}=\frac{(\partial_\omega q^{(+)}_{\mathbf{k}})^2}{q^{(+)}_{\mathbf{k}}}+\frac{(\partial_\omega q^{(-)}_{\mathbf{k}})^2} {q^{(-)}_{\mathbf{k}}} = \left( \partial_\omega q^{(\pm)}_{\mathbf{k}} \right)^2 \frac{q^{(+)}_{\mathbf{k}}+q^{(-)}_{\mathbf{k}}} {q^{(+)}_{\mathbf{k}} q^{(-)}_{\mathbf{k}}}.
\end{gather}
Here, the derivatives of the probabilities to measure $\ket{+}$ and $\ket{-}$ only differ by a sign; therefore, the squared derivatives are equal for both cases. We first rearrange Eq.~\eqref{eq:prob} as 
\begin{align}
q_{\pm,{\mathbf{k}}}
=& \left( \frac{\lambda_0}{2} \prod_{i=1}^n \lambda^{(i)}_{k_i}+\frac{\lambda_1}{2} \prod_{i=1}^n \lambda^{(i)}_{1-k_i} \right) \nonumber\\
&\pm  \left( \frac{\lambda_0}{2} \prod_{i=1}^n \lambda^{(i)}_{k_i}-\frac{\lambda_1}{2} \prod_{i=1}^n \lambda^{(i)}_{1-k_i} \right) e^{-(n+1)g t^{\alpha}} \cos[(n+1-2m_{\mathbf{k}})\omega t]
\end{align}
and substitute into Eq.~\eqref{smallf} to get
\begin{align}
\label{smallfintermediate}
f_{\mathbf{k}}=& t^2 e^{-2(n+1) g t^{\alpha}} (n+1-2m_{\mathbf{k}})^2 \sin^2[(n+1-2m_{\mathbf{k}})\omega t] \\ &\times \frac{\left( \lambda_0 \prod_{i=1}^n \lambda^{(i)}_{k_i}+\lambda_1 \prod_{i=1}^n \lambda^{(i)}_{1-k_i} \right) \left( \frac{\lambda_0}{2} \prod_{i=1}^n \lambda^{(i)}_{k_i}-\frac{\lambda_1}{2} \prod_{i=1}^n \lambda^{(i)}_{1-k_i} \right)^2 } {\left( \frac{\lambda_0}{2} \prod_{i=1}^n \lambda^{(i)}_{k_i}+\frac{\lambda_1}{2} \prod_{i=1}^n \lambda^{(i)}_{1-k_i} \right)^2 - \left( \frac{\lambda_0}{2} \prod_{i=1}^n \lambda^{(i)}_{k_i}-\frac{\lambda_1}{2} \prod_{i=1}^n \lambda^{(i)}_{1-k_i} \right)^2 \mathcal{E}} \nonumber,
\end{align}
where $\mathcal{E}=e^{-2(n+1) g t^{\alpha}} \cos^2[(n+1-2m_{\mathbf{k}})\omega t]$.

We want to look at the specific case where the Fisher information is maximised for any $\mathbf{k}$. If one assumes that $n$ is an even number (if it is odd, one of the RPQs can be discarded at the beginning of the protocol) and $\omega t =\frac{\pi}{2}$, then $(n+1-2m_{\mathbf{k}})$ is an odd number and the argument of the $\sin$ and $\cos$ functions has the form $\frac{\pi}{2}+ l \pi$, where $l$ is an integer. The $\sin$ function then gives one, and the $\cos$ function is equal to zero. This simplifies Eq.~\eqref{smallfintermediate} to
\begin{gather}
\label{smallffinal}
f_{\mathbf{k}}= t^2 e^{-2(n+1) g t^{\alpha}} \mathcal{P}_{\mathbf{k}} (n+1-2m_{\mathbf{k}})^2,
\end{gather}
where $\mathcal{P}_{\mathbf{k}}$ is defined in Eq.~\eqref{eq:Pofk}. To calculate the total CFI, one has to sum over all possible strings of measurement outcomes, characterized by $\mathbf{k}$:
\begin{gather}
\label{CFIeq}
\mathcal{F}^{\sharp}_C=\sum_{\mathbf{k}} f_{{\mathbf{k}}}= t_{\sharp}^2 e^{-2(n+1)g t_{\sharp}^{\alpha}} \sum_{\mathbf{k}} \mathcal{P}_{\mathbf{k}} (n+1-2m_{\mathbf{k}})^2.
\end{gather}
Here, $\mathcal{F}^{\sharp}_C$ is the time-optimised CFI, with optimal time for the correlated probes denoted by $t_{\sharp}$. The last equation is identical to Eq.~\eqref{QFIeq}. This means that the readout procedure introduced here achieves optimal sensitivity for estimating frequency. We no longer need to differentiate between QFI and CFI, but for concreteness we only use the abbreviation CFI from here on.

In Appendix~\ref{app:onequbmeas}, we derive the CFI for the case when only the SPQ is measured and all RPQs are discarded. Even then, we find that the CFI scales very much like in Eq.~\eqref{CFIeq}.

\subsection{Uncorrelated probe}
\label{sec:uncorr}

Before discussing our main results, let us compute the Fisher information for the case where the probe does not exploit correlations. Eventually, we want to compare the performance of the probe in the GHZ diagonal state to that of the uncorrelated state, i.e., the product state $\Omega_0 = \otimes_{i=0}^n \rho_i$.

For the $i$th qubit, the QFI can be easily computed to be $\mathcal{F}_Q^{(i)} = e^{-2 g t_\|} p_i^2$~\cite{arXiv:1003.1174}, which can be achieved with an $X$ basis measurement on each qubit, hence $\mathcal{F}_{Q}^{(i)} = \mathcal{F}_{C}^{(i)}$. In general, the run time for the uncorrelated probe will be different from that of the correlated probe; we denote the optimal run time for the former by $t_\|$.

As the Fisher information is additive, the total Fisher information for all qubits in $\Omega_0$ is
\begin{gather}\label{eq:ucFish}
\mathcal{F}_{C}^{\|} = \sum_i \mathcal{F}_C^{(i)} = e^{-2 g t_\|} (n+1) \braket{\mathbf{p}^2},
\end{gather}
where $\braket{\mathbf{p}^2} := \frac{1}{n+1}\sum_{i=0}^n p_i^2$ is the normalised length squared. This quantity is closely related to the average purity of all qubits, which has the form $\frac{1}{n+1} \sum_{i=0}^n \frac{1+p_i^2}{2} = \tfrac{1}{2}(1+\braket{\mathbf{p}^2})$.

\section{Features of the protocol}
\label{sec:features}

\subsection{`Zeno' scaling}
\label{sec:approx}

\begin{figure}
\centering
\includegraphics[width=0.65\textwidth]{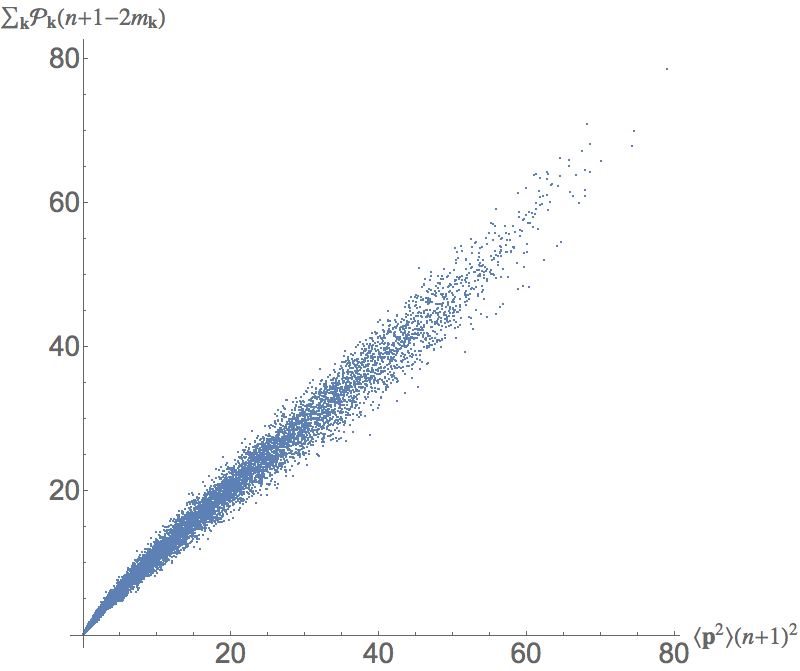}
\caption{\textbf{Approximating the CFI.} The relationship between $\sum_{\mathbf{k}} \mathcal{P}_{\mathbf{k}}  (n+1-2m_{\mathbf{k}})^2$ and $\braket{\mathbf{p}^2}(n+1)^2$. The plot shows linear behaviour with slope 1. The correlation function is $99.3 \%$. Calculations are made for $10^4$ randomly chosen vectors $\mathbf{p}$, with RPQ number $n \in [1,11]$ chosen randomly. The elements of the vector, $p_i \in [0,1]$, are chosen from the uniform probability distribution.}
\label{avpur}
\end{figure}

We have given an expression for the CFI in the last section that is rather complicated and does not shed much light on how well our probe performs. The complexity entirely lies in the final term of Eq.~\eqref{CFIeq}. It turns out that this term can be approximated, in terms of the vector $\mathbf{p}$ for any initial state, by $\sum_{\mathbf{k}} \mathcal{P}_{\mathbf{k}} (n+1-2m_{\mathbf{k}})^2 \approx \braket{\mathbf{p}^2}(n+1)^2$. This approximation holds remarkably well, as numerically evidenced in Fig.~\ref{avpur}. There, we plot $\sum_{\mathbf{k}} \mathcal{P}_{\mathbf{k}} (n+1-2m_{\mathbf{k}})^2$ against $\braket{\mathbf{p}^2}(n+1)^2$ for randomly chosen $1\leqslant n\leqslant 11$ (number of RPQs) and uniformly sampled $p_i \in [0,1]$. For $10^4$ realisations, the Pearson correlation coefficient is $99.3\%$. We have performed $\sim 10^6$ calculations for a larger range of the parameter $n \in [1,15]$, and the results hold. For clarity we show the reduced range results. This shows that for any vector $\mathbf{p}$ the CFI scales as $(n+1)^2$:
\begin{gather}\label{eq:CFIapprox}
\mathcal{F}_C^\sharp \approx t_{\sharp}^2 e^{-2(n+1)g t_{\sharp}^{\alpha}} 
\braket{\mathbf{p}^2} (n+1)^2.
\end{gather}

To compare the sensitivity of a correlated probe to that of an uncorrelated probe, we fix the total sensing time to $\mathcal{T}$, as well as the number of probes $n$, and compute the total Fisher information acquired. Each run of sensing with a correlated probe takes time $t_{\sharp}$, while sensing with an uncorrelated probe takes time $t_{\|}$. The total number of runs is given by $\mathcal{T}/t_{\sharp}$ and $\mathcal{T}/t_{\|}$ in the two cases respectively and the corresponding total CFIs are
\begin{align}
\label{CFItot}
F_C^\sharp =& \frac{\mathcal{T} \mathcal{F}_C^\sharp} {t_{\sharp}}  \approx  \mathcal{T} {t_{\sharp}} e^{-2(n+1) g t_{\sharp}^{\alpha}} \braket{\mathbf{p}^2}
(n+1)^2 \nonumber\\
\mbox{and} \qquad
F_C^{\|} =& \frac{\mathcal{T} \mathcal{F}_C^{\|}}{t_\|}=
\mathcal{T} t_\| e^{-2g t_\|^{\alpha}} \braket{\mathbf{p}^2} (n+1).
\end{align}
We then look at the case where the amount of noise in the system is the same after the experiment is concluded, i.e., we set  $\exp[-2g t_\|^{\alpha}] = \exp[-2(n+1) g t_{\sharp}^{\alpha}]$. Solving this for $t_{\sharp}$, we get $t_{\sharp} = t_{\|}/ \sqrt[\alpha]{n+1}$. Another way to motivate this choice is by finding the optimal values for $t_{\sharp}$ and $t_{\|}$ independently, and we find the same relationship between the two times.

By substituting this into the total Fisher information and taking the ratio of the two CFIs in the last equation, we get the quantum advantage
\begin{gather}
\label{eq:qadv}
\frac{F_C^\sharp}{F_C^{\|}} \sim 
\frac {(n+1)} {\sqrt[\alpha]{n+1}}.
\end{gather}
It is clear that the correlations in the probe give us an enhancement in the Fisher information that scales with the number of qubits. For the simplest time inhomogeneous noise model, where $\alpha=2$, one gets a CFI that scales as $(n+1)^{\frac{3}{2}}$. While for the semigroup case, with $\alpha=1$, no advantage is drawn from the correlated probe. In other words, we have shown that the Fisher information scales exactly as for the case where a pure probe state is used~\cite{PhysRevA.84.012103, PhysRevLett.109.233601, PhysRevA.92.010102}. This finding generalises the result of Ref.~\cite{arXiv:1003.1174}, where mixed state probes with $p_i=p$ for noiseless encoding were considered.

The sensing times $t_\|$ and $t_{\sharp}$ do not depend on the mixedness of the probe. The only difference is that we have an overhead of $\braket{\mathbf{p}^2}$, which can be overcome by repeating the experiment $\frac{1}{\braket{\mathbf{p}^2}}$ times. The correlations in the state of the probe lead to this enhancement for any $\braket{\mathbf{p}^2} \ne 0$. This includes highly mixed states that live in the separable ball~\cite{GurvitsBarnum2002} and thus cannot be entangled. Therefore, our work further highlights the importance of correlations beyond quantum entanglement in quantum metrology. This is one of our main results.

As mentioned before, in Appendix~\ref{app:onequbmeas} we consider the case where only the SPQ is measured and all RPQs are discarded. Even in this case, the `Zeno' scaling survives at the expense of an adaptive measurement protocol. We now examine two special cases of this general results.

\subsection{Uniform $(\star)$ and Tilted $(\$)$ probes}\label{app:cvc}

We now examine two special cases of our protocol. In the first case, we set all $p_i=p$ and call this the \emph{uniform} protocol. For the second case, we set $p_0=1$ and $p_i=0$~$\forall i\neq0$ and call it the \emph{tilted} protocol. Here all RPQs are maximally mixed and the SPQ is pure. These two protocols were previously studied for noiseless encoding in Refs.~\cite{arXiv:1003.1174} and \cite{PhysRevA.93.040304} respectively.

The initial probe state in the uniform protocol has a high degree of degeneracy, and its eigenvalues have the form $\left(\frac{1+p}{2} \right)^{n+1-m} \left(\frac{1-p}{2} \right)^{m}$. This simplifies the sum over the possible values of $\mathbf{k}$ in Eq.~\eqref{CFIeq}, and it can be rewritten as:
\begin{align}\nonumber
\mathcal{F}_{C \star}^\sharp =& t_{\sharp}^2 
e^{-2 (n+1)g t_{\sharp}^{\alpha}} \sum_{m_{\mathbf{k}}=0}^{n} {n \choose m_{\mathbf{k}}} \mathcal{P}_{m_{\mathbf{k}}} (n+1-2m_{\mathbf{k}})^2 \\
\geqslant& t_{\sharp}^2 
e^{-2 (n+1)g t_{\sharp}^{\alpha}} p^2 (n+1)^2.
\end{align}
The inequality was numerically shown to hold in Ref.~\cite{arXiv:1003.1174}, with the difference between the left and the right hand sides to be in relatively small. This is in perfect correspondence with our result in Eq.~\eqref{eq:CFIapprox}, since $\braket{\mathbf{p}^2}=p^2$ in this case.

In contrast, the state of the probe in the tilted protocol has eigenvalues $\{\frac{1}{2^n},0\}$, which simplifies the term $\mathcal{P}_{\mathbf{k}}$ in Eq.~\eqref{CFIeq} to $\frac{1}{2^n}$. There are exactly $n \choose m$ vectors $\mathbf{k}$ that have $m$ 1s and $n-m$ 0s; therefore, the CFI can be rewritten as:
\begin{gather}
\mathcal{F}_{C \$}^\sharp = t_{\sharp}^2 e^{-2 (n+1) g t^{\alpha}_{\sharp}} 
\frac{1}{2^n} \sum_{m_{\mathbf{k}}=0}^n {n \choose m_{\mathbf{k}}} (n+1-2m_{\mathbf{k}})^2
= t_{\sharp}^2 e^{-2 (n+1) g t^{\alpha}_{\sharp}} (n+1).
\end{gather}
This is in exact agreement with Eq.~\eqref{eq:CFIapprox}, for which $\braket{\mathbf{p}^2}=\frac{1}{n+1}$ here.

Now, consider the CFI for the tilted protocol for the case $\alpha=2$. We can simplify Eq.~\eqref{CFItot} to get $F_{C \$}^\sharp = \mathcal{T} t_\| \exp[-2 g t^2_\|] \sqrt{n+1}$. The scaling of $\sqrt{n+1}$ may give the impression that our quantum correlated probe performs worse than the standard quantum limit. However, if we were to use an uncorrelated probe consisting of one pure qubit and $n$ fully mixed qubits, after a time $\mathcal{T}$ the CFI would be $F_{C \$}^\| = \mathcal{T} t_\| \exp[-2 g t^2_\|]$. Thus, the correlations are still giving us an enhancement that grows with the number of qubits in the probe.

The constant terms of the CFIs for the uniform and the tilted protocols are identical. However, in terms of the probe size the former scales as $p^2 (n+1)^2$, and the latter scales as $n+1$. Thus, when $p \geqslant 1/\sqrt{n+1}$ the uniform protocol fares better and $p \leqslant 1/\sqrt{n+1}$ the tilted protocol fares better. This relation gives us a way to understand the trade-offs between the two extreme choices for the initial state of the probe. That is, when is it favourable for the coherence to be concentrated in a single qubit, and when is it favourable for it to be spread amongst all qubits.

Finally, in Ref.~\cite{PhysRevA.93.040304}, where the tilted protocol was originally introduced, the state preparation did not include the first \textsc{Cnot} gate (see Fig.~\ref{protocol}). This is of course  because the SPQ is pure and initially in state $\ket{0}$. The first \textsc{Cnot} gate would simply do nothing to the probe. However, it turns out that with the inclusion of this gate we need not care which qubit serves as the SPQ.

\subsection{Symmetry of $\mathbf{p}$}
\label{sec:symm}

We now show that CFI does not change under permutations of qubits or equivalently the elements of vector $\mathbf{p}$. This means our protocol does not depend on which qubit is used as SPQ. First, note that the term $\mathcal{P}_{\mathbf{k}}$ in Eq.~\eqref{CFIeq} is symmetric under permutations of the RPQs and thus changing their order does not change the value of CFI. Now, we consider swapping one RPQ with the SPQ. Without loss of generality, we consider a pair of vectors $\mathbf{p} :=(p_0,p_1,\dots,p_n)$ and $\mathbf{q} :=(p_1,p_0,\dots,p_n)$, where the first two entries have been switched.

We write the CFI given in Eq.~\eqref{CFIeq}, omitting the constant terms, for these two vectors: 
\begin{align}
 \mathcal{F}_{C,\mathbf{p}} \sim \sum_{\mathbf{k}} \frac{\left(\lambda^{(0)}_0 \lambda^{(1)}_{k_1} \prod_{i=2}^n \lambda^{(i)}_{k_i}-\lambda^{(0)}_1 \lambda^{(1)}_{1-k_1} \prod_{i=2}^n \lambda^{(i)}_{1-k_i} \right)^2}
{ \lambda^{(0)}_0 \lambda^{(1)}_{k_1} \prod_{i=2}^n \lambda^{(i)}_{k_i}+\lambda^{(0)}_1 \lambda^{(1)}_{1-k_1} \prod_{i=2}^n \lambda^{(i)}_{1-k_i}} (n+1-2m_{\mathbf{k}})^2, \\
\mathcal{F}_{C,\mathbf{q}} \sim 
\sum_{\mathbf{k}} \frac{\left(\lambda^{(1)}_0 \lambda^{(0)}_{k_0}  \prod_{i=2}^n \lambda^{(i)}_{k_i}-\lambda^{(1)}_1 \lambda^{(0)}_{1-k_0} \prod_{i=2}^n \lambda^{(i)}_{1-k_i} \right)^2}
{ \lambda^{(1)}_{0} \lambda^{(0)}_{k_0} \prod_{i=2}^n \lambda^{(i)}_{k_i}+ \lambda^{(1)}_{1} \lambda^{(0)}_{1-k_0} \prod_{i=2}^n \lambda^{(i)}_{1-k_i}} (n+1-2m_{\mathbf{k}})^2.\label{eq:qcfi}
\end{align}
Let's look at one term in the sum. The vector describing this term (excluding first RPQ) is $\mathbf{k'}=(k_2,\dots,k_n)$, with Hamming weight $m_{\mathbf{k'}}=\sum_{i=2}^n k_i$. We will show that for each term in the sum in $\mathcal{F}_{C,\mathbf{p}}$, there is an equivalent term in $\mathcal{F}_{C,\mathbf{q}}$. For a given vector $\mathbf{k'}$ and $\mathcal{F}_{C,\mathbf{p}}$ there are two options: $k_1=0$ or $k_1=1$.

When $k_1=0$ the corresponding term in $\mathcal{F}_{C,\mathbf{q}}$ will be term where $k_0=0$. For both $\mathbf{p}$ and $\mathbf{q}$ the Hamming weight of the vector $\mathbf{k}=(k_1,k_2,\dots,k_n)$ will be the same and equal to $m_{\mathbf{k}'}$, hence the multiplying factor $(n+1-2m_{\mathbf{k}'})^2$ will be the same.

When $k_1=1$ then the Hamming weight of vector $\mathbf{p}$ will be equal to $m_{\mathbf{k}'}+1$ and the multiplying factor will be $(n+1-2(m'_{\mathbf{k}}+1))^2=(n-1-2m_{\mathbf{k}})^2$. This term matches the term in $\mathcal{F}_{C,\mathbf{q}}$ with $k_0=1$ and vector $\mathbf{k''}=(1-k_2,\dots,1-k_n)$. This makes the numerator for Eq.~\eqref{eq:qcfi} $\left(\lambda^{(1)}_{0} \lambda^{(0)}_{1} \prod_{i=2}^n \lambda^{(i)}_{1-k_i} -\lambda^{(1)}_{1} \lambda^{(0)}_{0} \prod_{i=2}^n \lambda^{(i)}_{k_i} \right)^2$. The sign will be cancelled by squaring this term and the denominator will be the same as in $\mathcal{F}_{C,\mathbf{p}}$. The Hamming weight of the vector $\mathbf{q}$ in this case will be $n-m_{\mathbf{k}'}$, because the Hamming weight is calculated for vector $\mathbf{k}$ describing the product that goes with $\lambda^{(0)}_0$ eigenvalue of SPQ. The multiplying factor in this case will be the same: $(n+1-2(n-m_{\mathbf{k}'}))^2 = (-1)^2(n-1-2m_{\mathbf{k}'})^2$. 

We can repeat this argument for all terms in the sum. Also, since classical Fisher information is invariant under permutations of RPQs, the proof holds for switching SPQ with any of the RPQs.

The symmetry of our protocol does not simply stem from the fact that the prepared state of the probe is GHZ-diagonal. To illustrate this, consider the case where the first \textsc{Cnot} gate in our protocol is omitted, while keeping the remainder of the procedure identical (as in Fig.~\ref{protocol}). Using the arguments in Sec.~\ref{sec:prep}, it is easy to show that the prepared state in this case is also GHZ-diagonal. For concreteness we now work with three qubits and write down the CFI. Ignoring the constant terms that contain variables like $t$, $g$, etc., we find that CFI is proportional to the term with the sum, i.e., $p_0 (8 \lambda^{(1)}_0 \lambda^{(2)}_0 + 1)$. This quantity is clearly not symmetric under exchange of qubits. Therefore, the first \textsc{Cnot} gate, while seemingly useless, is responsible for the symmetry of our protocol. Moreover, as mentioned above, the first \textsc{Cnot} gate also leads to better sensitivity. To see this, consider the fact that for a small value of $p_0$ this CFI will be very small if the first \textsc{Cnot} gate is omitted. In contrast, the CFI for our protocol will be proportional to $3(p_0^2+p_1^2+p_2^2)$.

\subsection{Monotonicity}
\label{sec:mono}

\begin{figure}
    \centering
    \subfigure[]
    {\includegraphics[width=0.45\textwidth]{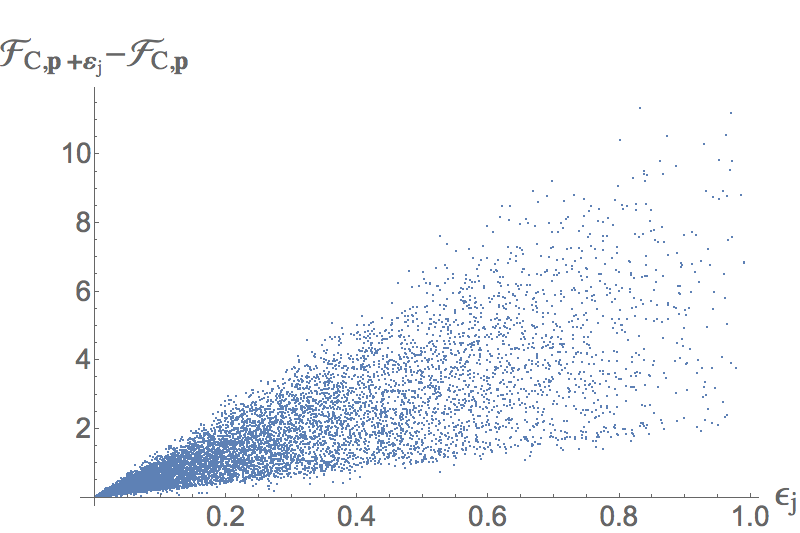}
    \label{CFI}}
    \subfigure[]
    {\includegraphics[width=0.45\textwidth]{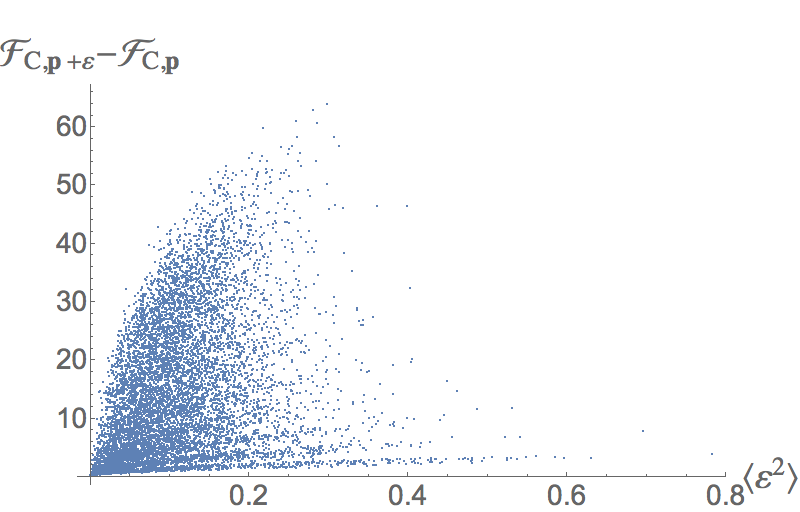}
    \label{CFIeps}}
    \caption{\textbf{Monotonicity of CFI in $\mathbf{p}$.} (a) The difference in CFIs $\mathcal{F}_{C,\mathbf{p}+\varepsilon_j}$ and $\mathcal{F}_{C\mathbf{p}}$ is plotted against $\epsilon_j$. Here $\varepsilon_j = (0,\dots, \epsilon_j, \dots,0)$ with constraints $\epsilon_j > 0$, and $p_j+\epsilon_j \leqslant 1$. The elements of $\mathbf{p}$ are sampled from the uniform probability distribution in the range $[0,1]$ and $0 \leqslant j \leqslant n$ is chosen randomly.
    (b) The difference in CFIs $\mathcal{F}_{C,\mathbf{p} + \varepsilon}$
    and $\mathcal{F}_{C,\mathbf{p}}$ against $\braket{\varepsilon^2} = \frac{\varepsilon \cdot \varepsilon}{n+1}$. Like $\mathbf{p}$, the elements of $\varepsilon$ are sampled from the uniform probability distribution in the range $[0,1]$ subject to constraint that all elements of $\mathbf{p} + \varepsilon$ are less than 1.
    Both calculations are made for $10^4$ realisations for randomly chosen $n \in [1,11]$, $t_{\sharp}=1$ and $g=0$. We did not observe a single event where the difference in the two CFIs was negative. We have performed $\sim 10^6$ calculations for a larger range of the parameter $n \in [1,15]$, and the results hold.}
\end{figure}

Our approximation for the CFI in Eq.~\eqref{eq:CFIapprox} is clearly monotonic in each of the elements of vector $\mathbf{p}$. We will now argue, again with numerical support, that the exact CFI in Eq.~\eqref{CFIeq} is indeed monotonic in this way. Since any permutation of the elements in vector $\mathbf{p}$ gives the same value of CFI, we can focus on changing just one parameter by adding a vector $\varepsilon_j= (0,0,\dots, \epsilon_j,\dots, 0)$, where $\epsilon_j>0$ and $p_j + \epsilon_j \leqslant 1$. We first show that CFI grows as any one parameter of vector $\mathbf{p}$ becomes larger: $\mathcal{F}_{C,\mathbf{p}+\varepsilon_j} \geqslant \mathcal{F}_{C,\mathbf{p}}$. 

We have tested the last inequality numerically for $10^4$ realisations; the results are presented in the Fig~\ref{CFI}. The plot shows that the difference in CFIs corresponding to randomly chosen $\mathbf{p}+\varepsilon_j$ and $\mathbf{p}$, subject to constrains from the last paragraph, is non-negative and grows with the size $\epsilon_j$. The implications of the numerics is that the CFI is monotonic in each elements of $\mathbf{p}$. Note that in this demonstration we are only interested whether the CFI grows and that the said difference is non-negative. The monotonicity stems from the monotonic nature of the sum in Eq.~\eqref{CFIeq}, and does not depend on $t_{\sharp}$ and $g$. For concreteness, we have taken $t_{\sharp}=1$ and $g =0$ for the purpose of the plot. These constants only contribute to a positive factor in the CFI and will be the same for any vector $\mathbf{p}$. Different values of $t_{\sharp}$ and $g$ will not change the sign of the difference, just its magnitude.

Building on this result, we can see that if $\epsilon_j>\epsilon_k>0$ than $\mathcal{F}_{C,\mathbf{p}+\varepsilon_j} \geqslant \mathcal{F}_{C,\mathbf{p}+\varepsilon_k}$, since we can change the order of the qubits and $\mathcal{F}_{C,\mathbf{p}+\varepsilon_j} \geqslant \mathcal{F}_{C,\mathbf{p}+\varepsilon'_j}$ for $\epsilon_j>\epsilon'_j=\epsilon_k$. Moreover, we can order the CFI for different vectors: $\mathcal{F}_{C,\mathbf{p}+\varepsilon_j+\varepsilon_k} \geqslant \mathcal{F}_{C,\mathbf{p}+\varepsilon_j} \geqslant \mathcal{F}_{C,\mathbf{p}+\varepsilon_k} \geqslant \mathcal{F}_{C,\mathbf{p}}$ for $\epsilon_j>\epsilon_k>0$. The Fig.~\ref{CFIeps} shows the difference between CFIs for arbitrary vector $\varepsilon = (\epsilon_0, \epsilon_1,\dots,\epsilon_n)$, where $\epsilon_j > 0$ for all $j$, as a function of normalized length squared of this vector $\braket{\varepsilon^2} = \frac{\varepsilon \cdot \varepsilon}{n+1}$. The difference grows with the $\braket{\varepsilon^2}$. The two figures present strong numerical evidence for monotnicity of CFI with respect to $\mathbf{p}$.

\textbf{Majorisation.} Finally, we present a negative result. Numerically we tested whether majorisation~\cite{Bengtsson2008} has an affect on the Fisher information. Specifically, we considered how the majorisation of the two normalised vectors $\mathbf{\hat{p}}$ and $\mathbf{\hat{q}}$ first. Here $\mathbf{\hat{p}}$ and $\mathbf{\hat{q}}$ describe two sets of $n+1$ qubits. We found that despite of majorisation the corresponding Fisher information do not have any hierarchy. We also considered the majorisation conditions for the spectrum of the initial state of the probe. In this case as well, we found no hierarchy for the corresponding Fisher information.

\section{Conclusions}
\label{sec:conclusion}

We have studied frequency estimation in the presence of noise with initially noisy probes. Our protocol begins by considering $n+1$ differently mixed qubits, all diagonal in the $Z$ basis. This model closely mimics an ensemble of atoms whose spins are aligned with a magnetic field. We then use one qubit to apply series of \textsc{Cnot} gates on the rest, resulting a GHZ-diagonal state for the probe onto which the desired parameter is encoded by allowing it to evolve freely. The free evolution is accompanied by unwanted time inhomogeneous dephasing noise. A measurement procedure for optimal readout, to be employed after the encoding, is specified.

For this protocol we have derived a simple formula that accurately approximates both the quantum and classical Fisher information in terms of the number of qubits in the probe and the average purity of the initial qubits. We have additionally shown that, despite the singling out of the SPQ, our protocol turns out to be symmetric under permutations of the qubits. This has practical implications, in that we are free to take any qubit as the control qubit. Finally, we have shown that the sensitivity of the probe is monotonic in the purity of the qubits, which clearly identifies a simple resource for a complex many-body probe.

Our most important finding is that the `Zeno' scaling is attainable independently of the mixedness of the initial probe state (except for in the trivial case when all RPQs are maximally mixed). While, the GHZ-diagonal state is entangled in general, it does become separable when the initial probe is (highly) mixed. In this case the state is confined to the separable ball~\cite{GurvitsBarnum2002}, but even here the correlated probe yields an enhancement over uncorrelated probes, provided the encoding noise is not described by a semigroup. The brings into question where lies the boundary between the quantum world and the classical world.

Similar enhancements, due to correlations, even for highly mixed states were reported in Ref.~\cite{arXiv:1003.1174} for phase estimation, and in Ref.~\cite{PhysRevLett.118.150601} for the charging power of quantum batteries. In all of these instances it remains unclear whether other quantum correlations, like quantum discord~\cite{rmp} or even quantum coherence~\cite{RevModPhys.89.041003}, are important in attaining the quantum enhanced scaling. In some sense in each of these studies, the enhancement seems to stem from collective coherence. This is most strongly evidenced by the fact that our protocol, like previous works, is highly sensitive to loss of even a single qubit. A detailed connection between quantum coherence and quantum metrology is explored in Ref.~\cite{giorda}.

Practically speaking, our results imply that quantum metrology may be robust against noise, provided that no qubits are lost during the sensing period. While we have conjectured that, for a given set of qubits, the protocol we have considered here leads to the best sensitivity for the probe, we are not able to prove this statement. Moreover, while the optimal pure states lead to the same sensitivity, this degeneracy may break for mixed states. Thus, the optimal preparation for the probe remains an open problem. Our analysis could be extended to a wider set of optimal states~\cite{PhysRevLett.115.170801} beyond GHZ-diagonal ones. Furthermore, in the article we have restricted ourselves to simple dephasing noise, it should be possible to generalise our findings for time-inhomogeneous, phase-covariant noise. It remains to be seen whether the $1/n^{\frac{5}{6}}$ scaling in Ref.~\cite{PhysRevLett.111.120401} or $1/n^{\frac{7}{8}}$ scaling in Ref.~\cite{haase2017fundamental} are also achievable with a noisy probe state. 

Finally, in a related manuscript~\cite{nottinghampaper}, an alternative figure of merit is considered, with energy, rather than total time, as the ``scarce resource.'' There, the uniform protocol is applied to an ensemble of thermal qubits in the presence of general phase-covariant noise, with the result that the notion of optimality changes considerably when one is concerned with conserving energy. Indeed, in this latter scenario, one finds that, even for time inhomogeneous noise, it is preferable to use small sized probes -- contrary to the case of limited total time, for which, as shown in this work, one can obtain an indefinite super-extensive (Zeno) scaling in the number of qubits.

{\bf Acknowledgments.} We are thankful to L. A. Correa and K. Macieszczak for helpful comments. We gratefully acknowledge funding from the Royal Society under the International Exchanges Programme (Grant No.~IE150570), the European Research Council under the StG GQCOP (Grant No.~637352), the Foundational Questions Institute (fqxi.org) under the Physics of the Observer Programme (Grant No.~FQXi-RFP-1601), and the COST Action MP1209: ``Thermodynamics in the quantum regime''. KM is Australian Research Council's  Future Fellow (FT160100073).

\appendix

\section{Measurement of SPQ only}
\label{app:onequbmeas}

\begin{figure}
    \centering
    \subfigure[]
    {\includegraphics[width=0.60\textwidth]{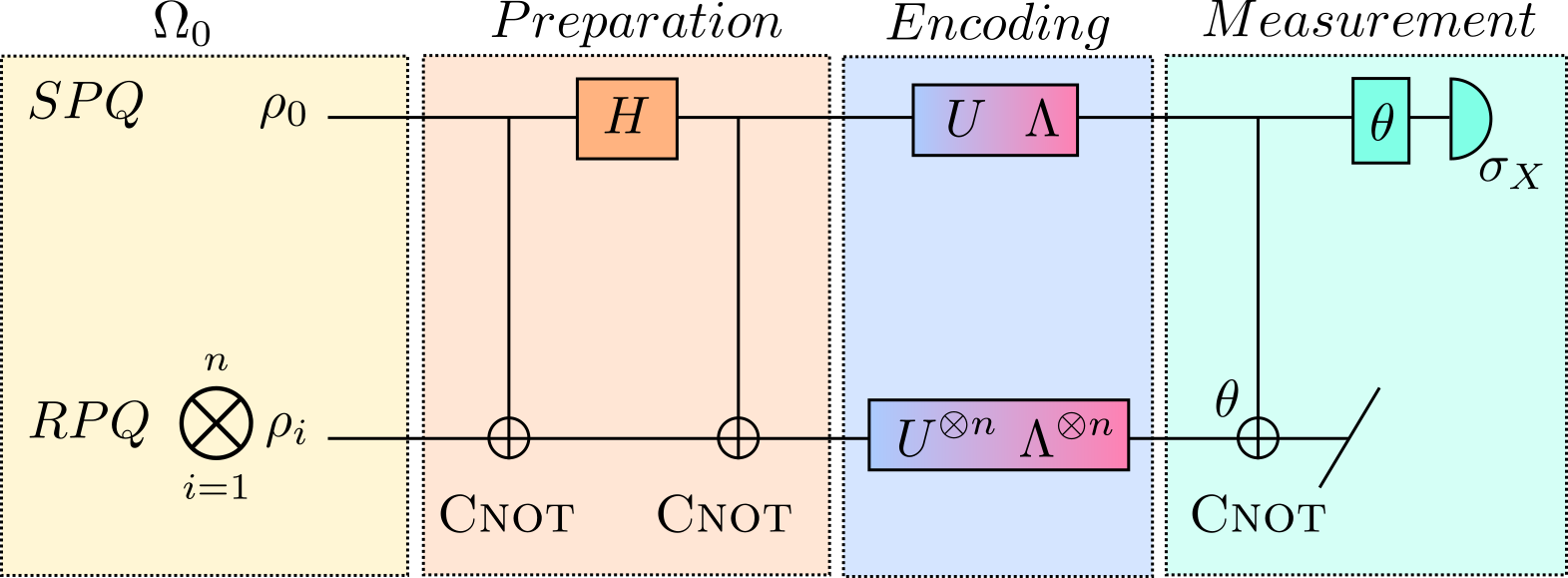}
    \label{protocolcap}}
    \quad\subfigure[]
    {\includegraphics[width=0.33 \textwidth]{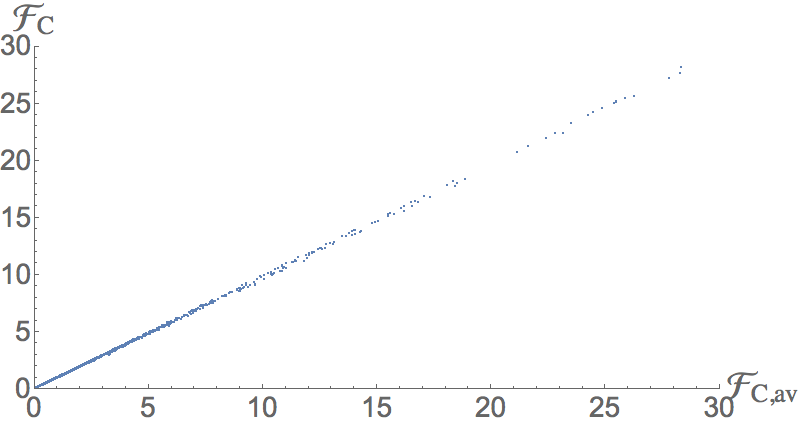}
    \label{CFIsingle}}
    \caption{\textbf{Single qubit readout.} (a) The protocol with measurement on SPQ only. Before discarding RPQs, a controlled operation is made to rotate them by an angle $\theta$ so that $\delta_\theta = \phi - \theta \approx 0$. (b) We approximate this CFI in terms averages of $\mathbf{p}$. The plot shows the CFI ($\mathcal{F}_C^{(1)}$) in Eq.~\eqref{CFIsingleeq} against its approximation in $\mathcal{F}_{C,av}^{(1)}$ in Eq.~\eqref{CFIsingleeqav}. The elements of $\mathbf{p}$ are sampled from the uniform probability distribution in range $[0,1]$. We have computed the two functions for $10^5$ random vectors $\mathbf{p}$ of random RPQ number $n \in [1,11]$. The plot is found to be linear with nearly unit slope, and the Pearson correlation coefficient is $99.99\%$.}
\end{figure}

We now show that it is possible to obtain near optimal CFI by only measuring the SPQ, while discarding all RPQs. This protocol is graphically illustrated in Fig.~\ref{protocolcap}. The main difference here is that, in place of the final \textsc{Cnot} gate, we do controlled rotation by an angle $\theta$ on the RPQs so that $\delta_\theta := \omega t - \theta \approx 0$. After the controlled operation, we discard the RPQs and measure the SPQ in the $X$ basis. The angle $\theta$ is our guess for the frequency $\omega$; thus, the sensitivity depends on our knowledge of $\omega$, which may be updated from run to run. The same strategy is employed in Ref.~\cite{PhysRevA.93.040304}.

The state just after discarding RPQs and before the measurement of the SPQ has the following form:
\begin{gather}
 \rho_{SPQ}= \frac{1}{2} 
  \left( \begin{array}{cc}
  1 & e^{-i \delta_\theta} e^{-(n+1)g t}
  \left(\lambda_0 \prod_{i=1}^n  x -\lambda_1 \prod_{i=1}^n  x^* \right) \\
   h.c. &  1
  \end{array} \right),
\end{gather}
where $x=\cos(\delta_\theta) - i p_i \sin(\delta_\theta)$. Next we measure the SPQ in the $X$ basis: 
\begin{align}
q_{\pm}=& \frac{1}{2}  \pm  \frac{1}{2} e^{-(n+1)g t} 
\nonumber\\ &\times
\Re \left[e^{-i \delta_\theta} \left(\lambda_0 \prod_{i=1}^n  [\cos(\delta_\theta) - i p_i \sin(\delta_\theta)] -\lambda_1 \prod_{i=1}^n  [\cos(\delta_\theta) + i p_i \sin(\delta_\theta)] \right)\right].
\end{align}
Since $\delta_\theta$ is small, we can expand the $\cos$ and $\sin$ functions and keep terms of the order of $\delta_\theta^2 t^2$ or smaller to get
\begin{align}
q_{\pm} =& \frac{1}{2} \pm \frac{1}{2} e^{-(n+1)g t} \nonumber\\
&\times \left( -\delta_\theta^2 t^2 \sum_{i=1}^{n} p_i +p_0 \left( 1- (n+1) \frac{\delta_\theta^2 t^2}{2} - \delta_\theta^2 t^2 \sum_{i=1}^n \sum_{i<j} p_i p_j \right) \right).
\end{align}

Next, we calculate the CFI according to Eq.~\eqref{cfieqgen}:
\begin{gather}
\label{CFIsingleeq}
\mathcal{F}_{C}^{(1)}=\frac{e^{-2(n+1)g t} \left( - 2\delta_\theta t^2 \sum_{i=1}^{n} p_i +p_0 \left( -(n+1) \delta_\theta t^2 - 2\delta_\theta t^2 \sum_{i=1}^n \sum_{i<j} p_i p_j \right) \right)^2}
{1 - e^{-2(n+1)g t} \left( -\delta_\theta^2 t^2 \sum_{i=1}^{n} p_i +p_0 \left( 1- (n+1) \frac{\delta_\theta^2 t^2}{2} - \delta_\theta^2 t^2 \sum_{i=1}^n \sum_{i<j} p_i p_j \right) \right)^2}.
\end{gather}

As before, we want to express the CFI in terms of an average of the elements of the vector $\mathbf{p}$. We approximate the sum over two parameters as $\sum_{i=1}^n \sum_{i<j} p_i p_j \approx \frac{n(n-1)}{2} \braket{\mathbf{p}}^2$, where $\braket{\mathbf{p}}:=\frac{\sum_{i=1}^{n} p_i}{n}$. With this, we can approximate the CFI as
\begin{gather}
\label{CFIsingleeqav}
\mathcal{F}_{C,av}^{(1)} \approx \frac{e^{-2(n+1)g t} \left(2\delta_\theta t^2 n \braket{\mathbf{p}} +p_0 \left((n+1) \delta_\theta t^2 + \delta_\theta t^2 n(n-1)\braket{\mathbf{p}}^2 \right) \right)^2}
{1 - e^{-2(n+1)g t} \left( -\delta_\theta^2 t^2 n \braket{\mathbf{p}} +p_0 \left( 1- (n+1) \frac{\delta_\theta^2 t^2}{2} - \delta_\theta^2 t^2 \frac{n(n-1)}{2}\braket{\mathbf{p}}^2 \right) \right)^2}.
\end{gather}
Fig.~\ref{CFIsingle} shows that this approximation holds very well, with $10^5$ random realisations of $\mathbf{p}$, whose elements are chosen from the uniform distribution in the range $[0,1]$, and random number of RPQs, $n \in [1,11]$. There is a linear trend between Eqs.~\ref{CFIsingleeq} and \ref{CFIsingleeqav} with nearly unit-slope and Pearson correlation coefficient of $99.99\%$.

We want to further simplify the expression for the CFI. Specifically, we want the denominator to be equal to one. For large values of $n$, this is the case when $\delta_\theta t \sim \sqrt{\frac{2}{n (1 + n \braket{\mathbf{p}}^2)}}$. The CFI has the following form:
\begin{gather}
\mathcal{F}_{C,av}^{(1)} \approx 2 e^{-2(n+1)g t} t^2 n^2 \ \frac{[2\braket{\mathbf{p}}^2+p_0 (1+n)]^2}{n(1 + n \braket{\mathbf{p}}^2)}.
\end{gather}
For large enough $n$, the ratio goes to 1 and we have the desired scaling for the CFI.

\bibliography{noisynoisy.bib}
\end{document}